\documentclass[manuscript]{acmart}

\AtBeginDocument{%
  \providecommand\BibTeX{{%
    \normalfont B\kern-0.5em{\scshape i\kern-0.25em b}\kern-0.8em\TeX}}}

\setcopyright{none}

\usepackage{xcolor}
\usepackage{csquotes}

\begin{document}

\title{Understanding the Digital News Consumption Experience During the COVID Pandemic}


\author{Mingrui ``Ray'' Zhang}
\affiliation{%
  \institution{The Information School, University of Washington}
  \city{Seattle}
  \country{USA}}
\email{mingrui@uw.edu}

\author{Ashley Boone}
\affiliation{%
  \institution{College of Computing, Georgia Institute of Technology}
  \city{Atlanta}
  \country{USA}}
\email{aboone34@gatech.edu}

\author{Sara M Behbakht}
\affiliation{%
  \institution{Human Centered Design \& Engineering, University of Washington}
  \city{Seattle}
  \country{USA}}
\email{sbehbakht2017@gmail.com}

\author{Alexis Hiniker}
\affiliation{%
  \institution{The Information School, University of Washington}
  \city{Seattle}
  \country{USA}}
\email{alexisr@uw.edu}


\begin{abstract}
During the COVID-19 pandemic, people sought information through digital news platforms. To investigate how to design these platforms to support users' needs in a crisis, we conducted a two-week diary study with 22 participants across the United States. \textcolor{black}{Participants' news-consumption experience followed two stages: in the \textbf{seeking} stage, participants increased their general consumption, motivated by three common informational needs---specifically, to find, understand and verify relevant news pieces. Participants then moved to the \textbf{sustaining} stage, and coping with the news emotionally became as important as their informational needs. We elicited design ideas from participants and used these to distill six themes for creating digital news platforms that provide better informational and emotional support during a crisis. Thus, we contribute, first, a model of users' needs over time with respect to engaging with crisis news, and second, example design concepts for supporting users' needs in each of these stages.}
\end{abstract}



\maketitle

\section{INTRODUCTION}
Designers have become increasingly influential in shaping people's news-consumption experiences. American adults are more likely to rely on interactive apps and websites for their news than on traditional forms of media (such as television, radio, and print formats) \cite{barthel_measuring_2020}, and 20\% of adults use social media as their primary news source \cite{shearer_social_nodate}. 
Unlike traditional news, digital news can be updated continuously, personalized~\cite{li_personalizenews_10}, and composed of interactive elements~\cite{hohman2020communicating}, creating numerous levers by which designers might influence users' experiences and behaviors when checking the news. This influence can be both positive and negative and can, for example, increase user engagement~\cite{Thurman2011Making}, improve relevance~\cite{Adar2017Personalog}, promote information overload~\cite{Josephine2018Toomuch}, or manufacture outrage via sensationalized content~\cite{vanderwicken_why_1995}. These divergent outcomes highlight the responsibility designers have to craft news platforms with care.


This responsibility is particularly conspicuous during times of crisis. During the uncertainty that arises in the wake of an unfolding disaster, people are dependent on the news to connect them with critical resources and information~\cite{stieglitz2018sense}, and they are more vulnerable to fear and anxiety~\cite{Hall2019TheAB}. Even in the best of times, information overload, online misinformation, and compulsive checking habits can undermine users' well-being~\cite{Song2017perceive}, and it is reasonable to predict that these threats to users' well-being may be exacerbated by a crisis. A large and diverse body of literature has investigated people's experiences with news media in many contexts, documenting, for example, where people get their news~\cite{shearer_social_nodate}, what news they find credible~\cite{Flintham2018falling}, what they share on social media in times of crisis \cite{letshate}, and how news about disasters spreads~\cite{Heverin2012use}. Prior work has also shown that specific patterns of news consumption during a crisis are linked to anxiety and depression~\cite{Bendau2020, excessive, Riehm2020associations}. Here, we build on this substantial foundation by: 1) eliciting user needs with respect to consuming news in times of crisis, and 2) identifying design principles for meeting these needs.

Specifically, we examine people's experiences using news platforms during a two-week window in the early stages of the COVID-19 crisis in the United States. Americans' consumption of news and current events increased 215\% in response to COVID-19~\cite{nielsen_navigating_2020}, signaling a clear demand for designs to support news consumption in this context. To better understand how designers can best support this scenario, we ask three research questions:

 \begin{enumerate}
    \item[] \textcolor{black}{RQ1. How, if at all, did people's patterns of engaging with digital news change during the COVID-19 crisis?}
    \item[] \textcolor{black}{RQ2. What user needs drove these patterns of engagement?}
    \item[] RQ3. What designs are likely to support these needs?
 \end{enumerate} \textcolor{black}{Our ultimate goal was to inform the design of interactive news platforms, such as news apps and websites (RQ3). To do so, we examined people's information-seeking experiences holistically (RQ1, RQ2), drawing on the construct of ``digital journalism'' \cite{digitaljournalbob, monica_how_nodate}---which includes news websites, alternative journalism sites, blogs, social media posts, news applications, and individual and group announcements---and looked at their engagement patterns broadly. 
} 

To answer these research questions, we conducted a two-week diary and interview study in April of 2020 with 22 participants, distributed geographically across the United States. All participants lived in areas with severe rates of COVID-19 disease activity at the time of the study, including New York, Michigan, and Washington. The study began with semi-structured interviews, in which we probed participants' news consumption experiences during the pandemic. Participants then completed a two-week diary study in which they recorded their daily news consumption behaviors, reflected on how these behaviors made them feel, and documented design ideas that occurred to them in response to their \textit{in situ} experiences. Drawing on themes that emerged across diary entries, the research team created prototype platform designs. Participants then provided feedback on these design concepts and shared reflections in a final exit interview.

\textcolor{black}{From our findings, we derived a formative conceptual model representing two stages of crisis news consumption: a \textit{seeking stage} and a \textit{sustaining stage}. In the initial \textit{seeking stage}, participants sought more information, invested increased time, and strove for greater variety of sources than they had in the past. They also relied more on local news. In this stage, \textit{informational needs}---the need to find, understand, and verify relevant news pieces efficiently---drove participants' news-engagement behaviors. As the crisis developed, people became exhausted and overwhelmed by the unrelenting inflow of information, especially negative news. Their news-engagement behaviors then shifted to a \textit{sustaining stage}, in which they tried to reduce their consumption, form intentional and bounded news-checking habits, and find a balance between ``being informed'' and ``getting too much.'' In this stage, \textit{emotional needs}---including the need to contain the time they spend with the news, cope with the negative sentiments it provoked, and connect with their family and friends around news items---became as important as informational ones.}

\textcolor{black}{Based on this conceptual model and on design feedback from participants, we created example interface components that designers might leverage to support the informational and emotional needs that arose for participants during this prolonged crisis. These components include, for example, a bounded sandbox for crisis-related news, visuals foregrounding the time a user spends reading the news, spatial separation of opinion pieces, and sentiment tags for articles.}
\textcolor{black}{In short, we contribute:
\begin{itemize}
    \item Empirical data about users' \textit{in situ} experiences with news media over time during the COVID-19 crisis which informed a taxonomy of common needs with respect to interactive news media. This in turn led to a conceptual model of digital news consumption during the COVID-19 crisis.
    \item Exemplar designs, developed iteratively with users, illustrating how interactive news platforms might support people in each of the dimensions documented in the model.\end{itemize}} Emerging literature (e.g.,~\cite{castriota2020national, excessive, Casero-Ripolles_2020}) has begun to document people's information seeking and engagement with mainstream media during the COVID-19 pandemic. This work has shown, for example, that people increased their news consumption during the pandemic~\cite{Casero-Ripolles_2020} and checked more local news~\cite{castriota2020national}. \textcolor{black}{However, this prior work provides only point-in-time snapshots of people's experiences with the news, rather than following the arc of their reflections and needs as the crisis wore on. It also does not attempt to draw design implications based on these experiences. Thus, we corroborate and expand on these past findings, and we contribute the model of user needs that emerged over time along with user-centered design guidance for creating interactive news media platforms to meet these needs.}
\section{Related Work}
 One prominent characteristic of current news consumption is the shift from traditional paper and television formats to interactive online sources, including social media platforms \cite{Bentley2019understanding, monica_how_nodate, nic2009a}. Interacting with news content (such as commenting and sharing a news post) has become part of the consumption experience and broadened the way people engage with news. Although these new platforms provide more timely and varied news compared to traditional media, users struggle to extract useful articles from the huge amount of available information. In this section, we review previous work on people's digital news consumption, negative experiences with online news, and news consumption and during times of crisis.

\subsection{Understanding Online News-Reading Behaviors}
Over the past decade, the internet has steadily become more entrenched as a first-class news source for many people. In 2012, the average American browsed five news pages per month \cite{Sharad2012who}, but by 2019, internet news consumption had increased significantly, and a series of surveys conducted by the Pew Research Center~\cite{shearer_social_nodate} showed that over 53\% of Americans got their news from online sources like websites and social media platforms. In 2019, Bentley et al. \cite{Bentley2019understanding} collected browser logs from 174 participants, finding that 23\% of all web-browsing sessions included views of news pages, and 23\% of this traffic originated from social media platforms. Most participants took in news from sources that presented differing perspectives.




A rich body of work has also investigated people's engagement patterns with online news in general. In one of the earliest investigations in this space, Aikat \cite{Debashis1998news} studied online news-reading behaviors on two news websites in 1998, describing patterns that continue to apply to today's news environment. Readers visited news websites more often during working hours than outside of work, and they usually skimmed headlines with short dwell times. More recently, Diakopoulos and Naaman \cite{Diakopoulos2011towards} studied the online news comments on sacbee.com, finding that people had different motivations for writing comments, including to share information, to express opinions, to debate or entertain, and to socially interact with others. Morgan et al. \cite{morgan2013is} investigated news-sharing behaviors on Twitter, and found that the more tweets a user sent, the more ideologically diverse the shared news items were. 

\textcolor{black}{Despite the public's eager engagement with digital news, it has also introduced new threats to users' mental wellbeing. For example, \textit{negativity bias} \cite{vaish_grossmann_woodward_2008} describes people's tendency to focus more on negative news than positive news. This bias, coupled with algorithms that prioritize content that attracts users' attention, has led to systems that are more likely to surface highly emotional content and make sensationalized content viral~\cite{essay78500}. 
Similarly, the huge amount of available news on digital platforms can lead to \textit{information overload}, wherein a taxing amount of information erodes mental health and inhibits decision-making \cite{buckland_information_2017}. When overloaded with news, people feel less confident about whether they have found the news they want \cite{PENTINA2014211}, tend to shut down cognitively, will deny the need to continue to consume news \cite{Linda2006the}, and will eventually cut back on news engagement \cite{PENTINA2014211}. Given the negative ways in which digital news platforms can impact users' wellbeing, we sought to understand not only how people used these platforms in a crisis, but also how the experience made them feel. Specifically, we found that as the crisis wore on, news-engagement became more emotionally taxing, leading people to create coping strategies, such forming new habits and seeking out positive content.}



\subsection{Digital News in Times of Crisis}
\textcolor{black}{Every crisis is different, and the way in which people experienced the COVID-19 pandemic may or may not align with people's experiences of other crises. According to the sociology of disaster research (p. 50 in \cite{Perry2005WhatIA}), crisis events can be categorized along several dimensions \cite{Olteanu2015what}, including the type of hazard (e.g., natural vs. human-induced), temporal development (instantaneous vs. progressive), and geographic spread (focalized vs. diffused). Previous research has found that information flow varies substantially across these different crisis types \cite{Olteanu2015what, Kanhabua2013understanding}. However, during crisis events, social media is often an important information source, and has been identified as a primary news resource \cite{Andrews2017}. The practice of using social media to actively engage with news information is also referred as ``citizen journalism'' \cite{atton2009alternative}, where public citizens ``\textit{[play] an active role in the process of collecting, reporting, analyzing, and disseminating news and information}'' \cite{bowman2003we}.}

\textcolor{black}{For instantaneous and natural crises such as earthquakes and floods, researchers have identified six main ways in which people leverage social media platforms such as Twitter and Weibo during a crisis, including: providing emotional support, identifying affected individuals, sharing information about donations and opportunities to volunteer, providing caution and advice, informing others about infrastructure and utilities, and reporting useful information \cite{Olteanu2015what, Qu2011microblogging}. Official organizations have also increasingly adopted social media platforms for making announcements in such crises \cite{Hughes2014online, DABNER201269}. In those types of crises, previous research demonstrated two roles of social media platforms: 1) because of the immediacy of social media, people tend to rely on information from these platforms to understand the situations and get resources. 2) because everyone can engage and participate, social media supports collaborative sense-making and the co-creation of knowledge during the time of crisis \cite{Stieglitz2017SensemakingAC, Palen2010avision, Macias2009blog, liu2008search}, such as \textit{information sharing and seeking, talking cure and understanding the "why"}\cite{Heverin2012use}. Inspired by these previous studies, we sought to understand participants' sense-making experience during their news consumption in COVID-19, and identified several informational needs during the procedure.}

\textcolor{black}{For progressive crises such as pandemics (including COVID-19), people also used social media and other news platforms to support their risk assessment and long-term decision making. For example, Gui et al. \cite{Gui2017managing} investigated how people used social media to make travel-related decisions during the Zika-virus outbreak. They found that local residents and previous travelers offered crucial resources that were absent from the news and other formal channels. People also began to form new consumption behaviors during the long-term crises, such as proactively looking for actionable news items \cite{howpetrun, Olteanu2015what} and shifting their news interests \cite{castriota2020national}. On the other hand, the long-lasting and intense news reports around the crisis also were negative in tone \cite{NBERw28110}, and could easily led to information overload \cite{info11080375, Austin2012HowAS} and mental health struggles like depression and anxiety \cite{GarfinThe2020, Bendau2020, Gaomental, chao2020media, Riehm2020associations, excessive} compared to the consumption in normal times \cite{Casero-Ripolles_2020}. In this project, we specifically looked into the negative effect caused by news consumption and how the experience revealed their emotional needs.}

\textcolor{black}{To date, this prior literature has primarily included surveys and interviews to understand people's experiences consuming COVID-19 news and information. We expand on this foundation, first, by capturing \textit{in situ} data about news consumption using diary methods, and derived a conceptual model that could be potentially applied to a broad spectrum of news consumption behaviors during the crisis times. We then draw upon this foundation to investigate how to design a positive experience for digital news consumption under such a crisis.}

\subsection{Designing Supportive Information Platforms in Times of Crisis}
\textcolor{black}{Given the problems and unique patterns of information consumption during crises, researchers have proposed designs and systems to improve the information platforms, including digital news and social media platforms to support the users' needs in the times of crisis. One major scope in this area is to combat the misinformation: prior work has examined people's trust in online news specifically in times of wildfire crisis, finding that family and friends were a more trusted information source than mass media \cite{Toddi2015what}, indicating that platforms should support for group and community-based information sharing mechanisms. Researchers also investigated the design elements that affect users' perceptions of the credibility of a news source \cite{Geeng2020Fake, Flintham2018falling, Hsueh2015leave, wobbrock2019isolating}. Specifically, Bhuiyan et al. \cite{Design2021trust} proposed design practices such as presenting appropriate details in news articles (e.g., showing the number and nature of corrections made to an article) to promote the transparency and trust.} 

\textcolor{black}{Previous research also focused on design mechanisms to deliver high quality news contents. For example, the functionalities such as comments and upvoting allowed the users to perform network gatekeeping and to organize and validate the breaking news feeds on Reddit \cite{upvote2017}. Zhang et al. \cite{mapping2021} reviewed 668 visualizations in COVID-19 news articles and synthesized themes that crisis visualizations should focus on. However, most of the previous work focused on designing a specific components (such as visualization, or commenting) for digital news platforms, and focused mainly on improving the information quality (such as clarity, credibility) of the news content. In this paper, via examining the dynamic behavior change of news consumption, we identified both \textit{informational} and \textit{emotional} needs during a crisis, and worked with participants together to present design ideas for the whole spectrum of digital news consumption.}



\section{Method}
To understand people's online news consumption experience during the COVID-19 pandemic, we conducted a two-week diary study with 22 participants geographically distributed across the United States. The diary study was preceded by an initial semi-structured interview and followed by a final design-feedback interview. \textcolor{black}{The study was approved by our university's Institutional Review Board.}


\subsection{Participants and Recruitment}
We conducted the study from April to June, 2020, a period that roughly corresponded to the first peak in COVID-19 disease activity in the United States \footnote{https://coronavirus.1point3acres.com/}. We aimed to \textcolor{black}{include participants with a mix of different backgrounds} (e.g., based on ethnicity, age, and location) to elicit a wide variety of news-reading practices. Thus, the research team posted recruitment advertisements on multiple city and state-specific subreddits and on several institutional mailing lists. Inclusion criteria were that the participant be over 18 years old, fluent in English, and use at least one technology device daily. We did not add specific criteria regarding news consumption, as we wanted to include participants with a wide range of news-reading habits.

We recruited 22 participants \textcolor{black}{(14 women and eight men, age 20-67, 64\% White, 23\% Asian and 14\% Black) living in six U.S. states (6 NY, 6 WA, 4 NJ, 2 MI, 2 CA and 2 PA) (for detailed demographic information, see Appendix \ref{sec:demo}).}

All participants lived in the places where COVID-19 disease activity was very high and ``stay-at-home'' orders had been been issued by the state government \footnote{https://www.cdc.gov/mmwr/volumes/69/wr/mm6935a2.htm}. All participants owned a smartphone and most used smartphones and computers regularly. \textcolor{black}{All participants gave their consent to participate in the study.}

\subsection{Procedure}
Each participant first attended in a one-hour semi-structured entrance interview remotely via Zoom, focusing on research questions R1 and R2. After explaining the procedure and the purpose of the study, we asked the participant about their technology use and news consumption behavior \textit{before} the pandemic, including how they accessed the news, how frequently they read the news, whether they engaged with the news or not (for example, by sharing or commenting on articles), and how they felt about these experiences before the pandemic. In keeping with the definition of digital journalism, we gave participants the freedom to bring up any ``online news'' experiences that were salient to them \cite{thatsnotnews}. We then asked questions about their online news consumption \textit{during} the pandemic, including changes to the news sources they depended on (if any), changes to their engagement (if any), their feelings about these changes (if any), and the challenges they encountered (if any). Finally, we asked them to pick one news-related application they used daily and demonstrate the steps they go through when checking news. A complete interview script is provided in Appendix~\ref{lb:interviewscript}. During this walk-through demonstration, participants were encouraged to describe how the features of the app affected their experience, either in a positive or negative way. Interviews were video-recorded, and participants were compensated with a US \$20 Amazon gift card for completing this portion of the study. 

After the entrance interview, all participants took part in the diary study, in which they completed a daily survey for two weeks. Survey questions are shown in Appendix \ref{sec:diary}. The questions focused on three topics: 1) the participant's news consumption, 2) how technology affected the participant's feelings about their news consumption, and 3) how they wanted to improve their news-consumption experience. \textcolor{black}{Participants could also attach a screenshot to document features they liked or disliked, to illustrate design ideas upon which they could elaborate}. To encourage high-quality answers and retention, we compensated participants only if they completed the survey more than five days per week (all participants met this threshold in our study). Participants received Amazon gift cards worth US \$20 for the first week and US \$40 for the second, with a gift card worth US \$10 as a bonus for each week where they provided particularly detailed answers. 

Based on the data from the entrance interviews and the diary study, we derived design themes for improving the crisis-time news consumption experience. We then made sketch interfaces demonstrating those themes and evaluated them via the exit interview of the diary study. Participants were reimbursed with a \$20 gift card for the one-hour interview. In all, participants could be compensated for a minimum \$100 and a maximum \$120 if completing all the three stages.

\subsection{Data Analysis}
\label{sec:data}

In all, we collected data from 22 interview sessions and 275 diary entries. We transcribed all interviews and coded transcripts and diary responses using an inductive analysis approach \cite{Corbin1990GroundedTR}. For each participant, we also examined changes in habits and in feelings throughout the two weeks. \textcolor{black}{After individually reviewing all data, three members of the research team met collaboratively online to discuss themes that emerged. The team refined these themes by reviewing specific examples together, leading to an initial codebook. Each entry in the codebook included the category name, an explanation, a list of defined subcategories, and example quotes for each subcategory. The first author then coded the data using this preliminary codebook and refined its categories. The final codebook contained four categories (\textit{behaviors, feelings, coping strategies, design-ideas}), and 41 nested subcategories (e.g., under \textit{behavior} were \textit{increase in consumption time}, \textit{increase in resource diversity}).} 

\textcolor{black}{We conducted a separate analysis of the design ideas we collected from the participants through entrance interviews and diary entries. We reviewed all designs and then constructed
affinity diagrams \cite{HOLTZBLATT2005159} to organize similar designs into coherent themes. The diagrams were created via Miro, an online collaboration platform. The final diagram contained 36 subcategories organized into six themes (e.g., \textit{filtering and customization}, \textit{organizing  information}, \textit{social factors}).}
\textcolor{black}{We used these themes to create interface sketches, which we presented to participants for feedback in the exit interview (see Section~\ref{sec:imaginary}). We conducted exit interviews with 17 participants, and their feedback was also coded via the thematic analysis and affinity diagramming.}

\section{A Conceptual Model for Crisis News Consumption}
\begin{figure}[h]
\centering
\includegraphics[width=\textwidth]{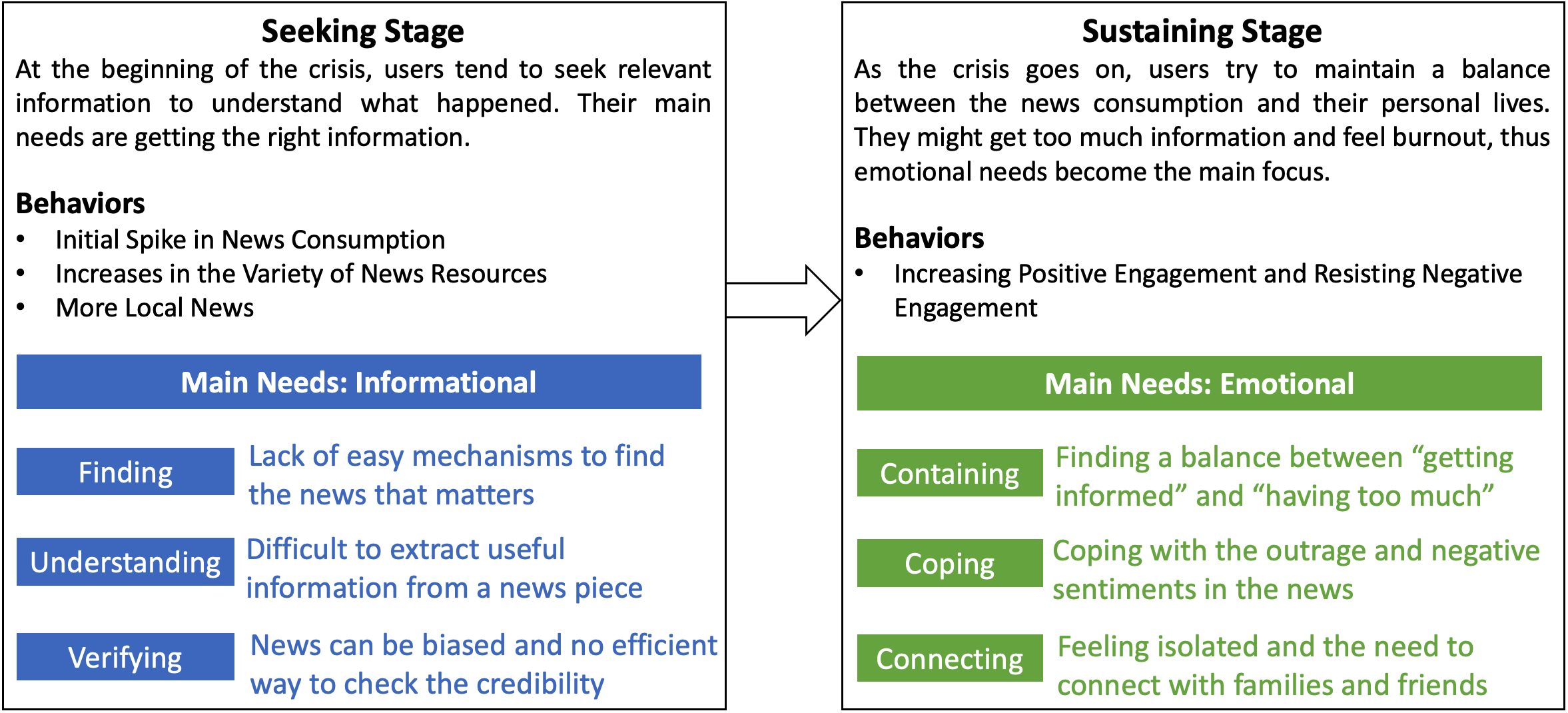}
\caption{\textcolor{black}{Two stages of digital news consumption during the COVID-19 crisis, characterized by consumption behaviors and the needs driving these behaviors}}
\label{fig:stages}
\end{figure}

\textcolor{black}{Across the themes that emerged from our data, we found that participants' news consumption clustered into two different temporal stages (see Figure~\ref{fig:stages}). At the beginning of the crisis, participants expressed a common set of behaviors and needs that reflect what we term the \textit{Seeking Stage}. In this stage, participants increased their consumption time and the diversity of the sources they sought out, and they focused on local news and sought out actionable activities. As time went on, participants shifted into what we term the \textit{Sustaining Stage}, wherein new emotional needs emerged as a result of the prolonged crisis and participants' new news-consumption behaviors. Feeling saturated with an excess of crisis-related news (news that was often negative, irritating, or sad), participants said they began to look for positive news pieces for emotional relief. In this stage, emotional needs became the main focus, as participants wanted to feel connected to others and in control of their information consumption.  In the following sections, we present the data that informs this model and, specifically, the 1) behaviors (Section~\ref{sec:behaviors} change and the main needs in two stages in more depth.}

\section{News-Consumption Behaviors During the Pandemic}
\label{sec:behaviors}
\textcolor{black}{We first describe how our participants changed their news consumption behaviors during the pandemic. Along with these changes, participants described what they valued and what they found challenging when engaging with news platforms, and we describe these user needs in Section~\ref{sec:needs}.}

\textcolor{black}{In the \textit{Seeking Stage}, participants reported three common ways in which their news consumption changed in response to the pandemic, specifically, \textit{an initial spike in consumption time, increase in the variety of news sources and more local news}; however, after having enough news consumption in the first stage, participants moved to the \textit{Sustaining Stage}. The most salient change in this stage was {increasing positive engagement and resisting negative engagement}. We elaborate on each of these themes below.}


\subsection{Seeking Stage: An Initial Spike in News Consumption}
All participants mentioned increasing their time with digital news as the pandemic took hold. Due to stay-at-home orders, people spent more time in their homes and on technology devices, giving them greater opportunity to check the news just as the crisis became urgent. At the beginning of the pandemic, participants mentioned seeking information actively, in contrast to their prior approach to engaging with the news passively. \textcolor{black}{One reason for this change was that at the start of the crisis, they wanted to learn as much as possible about this completely novel situation created by a new pathogen. For example, P3 set a dashboard of COVID cases in the U.S. as her homepage so that she could track the numbers all the time. P12 explained that she had become more likely to ``\textit{seek out}'' content proactively: ``\textit{before COVID my husband and I, at night, we would watch, like, a TV show on Netflix. But now when he's done with work, you know, he wants to see what is going on in the world.}''}  

\textcolor{black}{People also explained that they began actively checking the news to find resources and plan their personal life. For example, P16 described ``\textit{planing my life around it (the news)}'' and started to hoard toilet paper when he saw the related tweets. P11 started to check news about the flights and hotels, as she had future travel plans. P5 constantly checked for pandemic-related updates on local government websites to ``\textit{plan out, like, if I have to go to the grocery store and it's in a certain city}''.}

\textcolor{black}{They also described being more likely to share content, such as sharing resources with older family members, sharing official information to combat the spread of misinformation, and sharing articles to raise awareness about the pandemic. P4 mentioned reading news and sharing resources with her older relatives who did not use technology:
``\textit{I was also checking a lot of the news websites because I wasn't just purchasing for myself. I was also trying to make some purchases for my relatives in another state. Because they are old, they don't know how to order anything online.''}}

\subsection{Seeking Stage: Increases in the Variety of News Sources}
We also found that participants increased the variety of their news sources. At the time of the study, all participants were using more than one platform to access news about COVID-19. According to diary responses, 18 participants used news websites and services, such as The New York Times and Apple News, and 18 participants used social media platforms such as Reddit and Twitter. Although text was the most common format for the news participants consumed, nine participants also listened to podcasts, and eight participants watched online videos. For example, P10 mentioned that she started to listen to podcasts when she was exercising, as it did not require her full attention. P18 started watching videos on YouTube because she could ``\textit{see what's going on and feel the emotions.}'' 

Participants also increased their use of social media, especially Twitter: eight participants mentioned starting to use Twitter to access the news or using it more often than they had in the past. Participants explained that Twitter offered a convenient way to access local news and updates from official organizations in real time:

\begin{displayquote}
\textit{I also started to use Twitter to follow a whole bunch of local government accounts to get local news. I never used Twitter for news before, but now on Twitter, like, I followed the Seattle Times and the local Bellevue newspaper. --- P11}
\end{displayquote} 

Six participants also started to subscribe to multiple media platforms with different perspectives. For example, P13 began reading news from sources that included perspectives from different regions of the world, including CNN and the Guardian, to better understand how other countries were reporting the pandemic. P12 explained that she would cycle through ``\textit{all different [news] sites back and forth, seeing if they contradict each other.}'' P18 also encouraged others to do the same by sharing news offering what she perceived to be a balanced perspective:

\begin{displayquote}
\textit{I have different ways to get news because people from different countries have their own opinions or perspectives on the same thing\ldots For example, Asian people wear masks and think it's helpful. But in the beginning, people in the western countries may be on the opposite of opinion\ldots And I do feel if you just have one side of information you're more likely to be biased\ldots I had talked to my co-workers about my experience in China. They don't really believe people are locked in their homes, but now I'm not sure if they changed their views. --- P18}
\end{displayquote}


\subsection{Seeking Stage: More Local News}
While participants had previously watched more national and state-wide news, they became focused on what happened in their neighborhood, \textit{i.e.} local news once the pandemic took hold. For example, in response to the pandemic, several participants subscribed local subreddits and followed local organizations on Twitter. P9 mentioned following a local subreddit which she had not been interested in before:

\begin{displayquote}
\textit{I've definitely added the New Jersey one. Before, I didn't follow anything there because all they did was about silly things. Now it's more like focus on the new rules that are going into place, like the the ordinances and how people feel about. --- P9}
\end{displayquote} 

Some participants also began paying attention to news sources in their family members' cities to understand what was going on in local context of their loved ones. For example, P17 mentioned checking the news of her children's location:

\begin{displayquote}
\textit{Well, our state has a ``April 30 stay in place'' order. What does Wisconsin have, what does Boston have? Mainly the places that I have relatives in. I'll see check what's going on in Seattle, and I have a daughter in Virginia, so I'm kind of focused on where my close relatives live to see what's going on there. --- P17}
\end{displayquote}

Other participants said that they started checking the news from ``\textit{Governor Cuomo}'' (P15), ``\textit{certain countries like Italy because of friends}'' (P13), ``\textit{state websites like Way County}'' (P5) and numerous other locally specific sources.

\subsection{Sustaining Stage: Increasing Positive Engagement and Resisting Negative Engagement}
\textcolor{black}{All participants said that after an initial spike in news consumption to learn about the virus and ways to prevent it, they intentionally began to take steps to manage their news consumption and to engage with the news in more bounded, more positive, and more interpersonal ways. For example, three participants mentioned volunteering to help others by answering their questions on social platforms. P10, an attorney who did a lot of research on employment policies during the COVID-19 crisis, started to answer questions about unemployment on Reddit. P1 provided answers on Reddit to feel connected with others.}

\textcolor{black}{In other instances, people said they made changes to cut down on or manage their usage.} Overall, sixteen participants reduced the time they spent consuming news (i.e., their second week average time was less than the first week), three participants maintained similar levels of consumption time, and three increased their consumption time. \textcolor{black}{Similarly, three participants mentioned that they made an intentional shift to form regular news-checking habits instead of endlessly scrolling, as a way to stay ``in control'' (P7) of their news consumption. And P11 started to listen to more broadcasts when cooking and exercising, so that she could focus on other activities while also getting informed.}


\textcolor{black}{Participants also shifted the content they consumed, more actively seeking positive content as time went on. P10 and P17 both mentioned that they started to look for positive news, even they were interested in ``depressing news'' before COVID. P3 and P7 started to consume more ``research'' style news, such as lectures and talks about COVID-19, that were less sensationalized and emotionally charged.}




At the same time, participants actively tried to reduce patterns of engaging with the news that affected them negatively. Participants mentioned about getting information overload, and started to decrease their consumption intentionally. For example, six participants mentioned that they tried to avoid political and opinion-based news. P16 explained in one diary entry that he found he had ``\textit{spent entirely too much time arguing about COVID-19 and debating with people today},'' something he said would have been unusual before the pandemic. In response, he tried to consume less news by not using his computer. P5 also found that she was irritated by her friends on Facebook sharing opinions she did not agree with. At first, she tried to post long and thoughtful comments, finding that she `` \textit{felt more dread today because I encountered more resistance-type mentality}.'' As the pandemic went on, she reflected on her engagement and started to step back from the comments and posts. At the end of the diary study, she wrote ``\textit{I felt okay because my engagement was lighter and more carefree. So long as I don’t take others opinions seriously, I won’t be as affected}.''





\section{News-Consumption Needs During the Pandemic}
\label{sec:needs}

Participants reported a variety of benefits and challenges arising from their experience with digital news. These illuminated several underlying needs, which clustered into two broad categories: \textit{informational needs} and \textit{emotional needs}. \textcolor{black}{Specifically, \textit{informational needs} were the main drivers of users' behaviors in the \textit{Seeking Stage}, as the crisis began and participants tried to form a comprehensive understanding of the crisis and how they might respond. As time wore on, and participants continued to face an exhausting stream of largely negative news, their \textit{emotional needs} began to emerge and, along with information needs, began a primary driver of their behaviors in the \textit{Sustaining Stage}.}

\subsection{Seeking Stage: Informational Needs}
Participants wanted to efficiently get information that mattered to them in the \textit{Seeking Stage}. Although participants said they had more time to spend checking the news during the lockdown, the explosion of information and constant updates created enormous barriers to finding and digesting useful and truthful information efficiently. P12 mentioned that she kept the news playing on her television all day so she would not miss anything. As participants confronted this flood of information through digital platforms, they consistently struggled with 1) \textit{finding}, 2) \textit{understanding}, and 3) \textit{verifying} the information they encountered.

\subsubsection{\textbf{Finding}} \textcolor{black}{Fourteen} participants explained that they struggled to sift through the sea of possible articles, and they found that it was not always easy to find the news that mattered to them. As they used multiple platforms for digital news, they frequently encountered redundant content. For example, although P17 intentionally cycled through different websites to ensure she engaged with content from different perspectives, she found that achieving this well-informed view required wading through repetitive coverage:

\begin{displayquote}
\textit{Most of the sites have similar news stories like they might be in a headline position on one site and down lower in another website. I'm always hoping to find some information on one site that's not anywhere else. --- P17}
\end{displayquote}

Participants also struggled to find content that was useful and actionable, something many said they were seeking. They explained that they felt informed after reading the news, but they also felt helpless, as they did not know what to do about the upsetting information they encountered. To combat this, 
participants continued to look for news, but curated what they saw; for example, P16 muted many news media accounts on Twitter because they only reported the death counts without providing useful resources.
Some participants directly went to government websites trying to find relevant resources they were looking for. However, they also found that these websites were ``\textit{clumsy}'' and not updated frequently. For example, P10 described checking for policies on paid sick leave on a government website, saying:

\begin{displayquote}
\textit{So the other day I was looking for information about New York state mandated COVID 19 paid sick leave. I was trying to figure out what the employer needed to [do] \ldots And there's a little tiny piece here about paid sick leave, but the majority of this is actually about paid family leave. It doesn't talk about the COVID 19 sick leave\ldots about the new legislation and it doesn't give me any guidance\ldots so this was a really, really frustrating website. --- P10}
\end{displayquote}

When participants turned to social media for news, they continued to struggle to sift through content and prioritize information they were interested in. They explained that the lack of coherent organization and categorization of posts made it difficult to zero in on particular topics or make sense of the landscape of information holistically. P11 followed many accounts on Twitter, but she found that it was hard to track them, as there was no way to prioritize and organize the feed by topics. 



Collectively, these examples illustrate participants' common desire to easily separate the information they are looking for from everything else. The difficulty they encountered in doing so led them to spend a frustrating amount of time digging online, and in some instances, to avoid consuming news in a noisy environment.

\subsubsection{\textbf{Understanding}} Even when participants found a news article they thought would be useful, they \textcolor{black}{(n=11)} often struggled to understand its contents. For news on Twitter, content was limited to 280 characters. Without further context, this information was often incomplete or even misleading. 
On the other hand, news pieces containing too much content also caused participants frustration, because the useful points were buried in a huge amount of information. P21 mentioned that he ``\textit{hated long articles}'' because he lacked patience; P7 mentioned that he preferred clips rather than the whole segment when watching video news. 

Participants frequently cited statistics and diagrams as elements that improved their understanding. Many participants mentioned the usefulness of statistics, as it gave them direct access to the phenomenon being described in the article. Diagrams, especially geographic ones were preferred by many participants, although sometimes their meanings could be challenging to understand. For example, P18 described the difficulty of understanding a scientific diagram saying:
``\textit{I feel it's just like a long (diagram) and it stays the same there, and the value versus log---I don't even understand, what does that mean.}''

\subsubsection{\textbf{Verifying}} As information about the pandemic exploded, most participants \textcolor{black}{(n=19)} said assessing the credibility of the news was vital but sometimes difficult. Several participants expressed concern about accessing credible information on social media platforms, as there were too many opinions and too few facts:

\begin{displayquote}
\textit{I think that it (Twitter) can get into a pretty toxic space, especially, sharing things without actually reading it and just going off the headlines... It's interesting to see where people are -- especially politicians -- like if they're pro or against something, But I don't find it really useful in terms of its actual content. --- P3}
\end{displayquote}


Participants also realized that even credible information could be biased. For example, P12 mentioned that she needed to ``\textit{switch up the channels here and there}'' to get a balanced view of the news, \textcolor{black}{but different sources also had different types of news ``\textit{they chose to put forward},'' which could lead to bias.}


Three participants also utilized the comment section as a tool to validate the credibility of an article. P1 mentioned that someone might post the original source behind a news piece, which helped her to validate the content. However, participants like P17 found that the comments were not useful, because most of them were opinions: ``\textit{They're just people's opinions. Some of them are just very inflammatory, and I don't put any credence in them. Although a couple of them make you think.}''

\subsection{Sustaining Stage: Emotional Needs}
\textcolor{black}{Although reading the news makes one feel informed, it can also negatively affect mental health. This challenge problem became more salient for participants in the \textit{Sustaining Stage}.} We found that almost all participants expressed frustration and a sense of feeling overwhelming by the news, especially in the later part of the diary responses. Participants surfaced three broad categories of emotional needs, which we labeled: 1) containing 2) coping and 3) connecting.

\subsubsection{\textbf{Containing}} \textcolor{black}{Twelve} participants expressed a sense of ``\textit{being a slave to the news},'' saying they could not resist checking continually. For example, P16 spent most of his time refreshing and searching for news about COVID, because he was afraid of missing something important. Although aware of the anxiety caused by his consumption habits, he said that checking news was ``\textit{the only way I have control over to keep informed}''.


Even those who did not spend much time consuming news felt overwhelmed at the beginning of the crisis, because everyone was talking about the event. P20 mentioned that she ``\textit{did not even have to go looking for the information, just because every single person on the Instagram was posting stuff about the COVID}.'' 



The feeling of being overwhelmed gradually decreased for some participants as the pandemic wore on, and most participants decreased their news consumption relative to the beginning of the crisis. However, diary responses show that many participants were still trying to find a balance between ``\textit{getting informed}'' and ``\textit{consuming too much}.''

\subsubsection{\textbf{Coping}} Not only did the quantity of news stories explode at the start of the pandemic, a large portion of the news was also negative. Moreover, unlike negative news stories before the pandemic, COVID-19 was of direct personal relevance to participants and their loved ones. \textcolor{black}{Fourteen participants reported struggling with the overall negative sentiment of the news.} P20 described being more emotionally affected by the news than before, saying:

\begin{displayquote}
\textit{Now when I'm looking at the news, it's more emotional. And I would be stressing more if I'm reading a title that is shocking. But beforehand I knew nothing bad is going on in the world necessarily. It was not a stressful thing, to be honest, to read the news every day. --- P20}
\end{displayquote}

When participants realized that the news came with an emotional tax, they often wanted to take a step back. But by then, they found that COVID has already penetrated every aspect of their lives. For example, P7 used to listen to fitness and cigar podcasts, and they all began to talk about the pandemic. P9 was exposed to COVID-related news on her Instagram feed, which left her feeling frustrated, because she used the platform for entertainment. As a result of encountering negative crisis-related news at every turn,  participants wrote that they felt, for example, ``\textit{exhausted}'' (P4) and ``\textit{numb}'' (P13).


\subsubsection{\textbf{Connecting}} \label{sec:connect} Because of the ``stay-at-home'' orders, many participants spent their time at home without physically interacting with others. \textcolor{black}{Seventeen participants mentioned that reading news helped them feel connected with the outside world.} When these attempts were successful, they were emotionally rejuvenating: P5 mentioned that reading news articles helped her establish common topics with her friends when talking over the phone. P10 mentioned feeling ``\textit{isolated}'' and that reading and sharing news was a coping strategy to stay virtually connected with her family.


However, in other instances, participants attempts to connect with others failed. P13 tried to engage with his relatives who did not care about the virus by sharing the news, but ``\textit{it just doesn't work. So it's probably best to just let it come and not respond back because it just fuels the fire a little bit more.}'' When participants encountered the situations like this, they usually ended up passively decreasing or stopping the interaction with the members. 




Collectively, these findings illustrate a variety of struggles with respect to news consumption during the COVID-19 crisis. Participants were more motivated than usual to look for personally relevant news and resources, and they felt frustrated when this information was not accessible. Yet, due to the flood of news about the pandemic, they often found they consumed both too much news and not enough. Emotionally, they found it exhausting to check the news, even though they sought out this information voluntarily. And when they tried to shut it out, they found it was all around them. After identifying these needs, we sought to explore designs to improve these experiences, as we describe in the next section.

\section{Designing a Better Digital News Consumption Experience}
\label{sec:design}

\begin{figure}[h]
\centering
\includegraphics[width=0.8\textwidth]{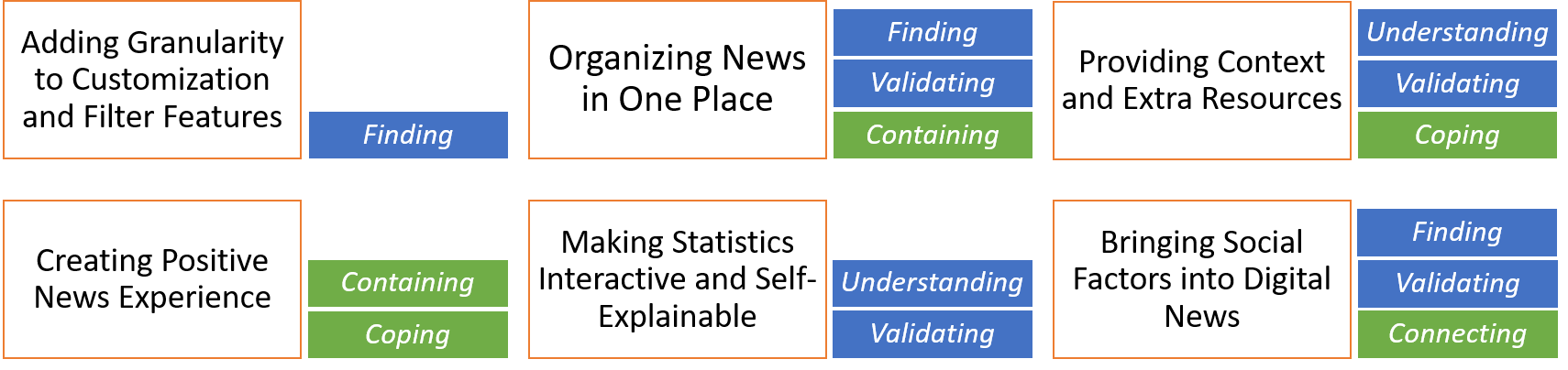}
\caption{Six designs themes resulted from the data. Each design is mapped to different needs.}
\label{fig:design}
\end{figure}

\textcolor{black}{After modelling participants' consumption behavior and identifying the needs driving these behaviors, we generated design ideas that digital news platforms could apply to support these needs.} The design ideas were generated via a user-centered approach: during the diary survey, we asked participants to reflect on their daily engagement with the news and to write down ideas for improving these experiences. We analyzed the data using affinity diagrams described in section \ref{sec:data}, which resulted in six design themes: 1) customization and filtering 2) organizing news in one place 3) providing context and resources 4) creating positive news experience 5) making statistics interactive and self-explanatory 6) bringing social factors into digital news. Each theme is described below, and all are shown in Fig \ref{fig:design}.

\subsection{Customization and Filtering}
Providing more granularity of news sorting and filtering mechanisms helps the user find relevant information efficiently. For example, participants suggested the following design concepts, all of which call more granular customization and filtering. All examples address the informational need \textit{finding}.

\textit{\textbf{Sorting Feeds by Popularity.}} Five participants mentioned using Reddit for getting news as it aggregated the news and sorted the information based on human interest. \textcolor{black}{This crowd-sourced approach of sorting by popularity reflects the concept of \textit{collective attention} \cite{huberman2008crowdsourcing}: helping the user grasp important information with limited attention resources by leveraging the input of the crowd.}
For example, P8 mentioned that by skimming through popular posts he can quickly get a sense of what is happening. Providing alternative sorting options, such as by views or comments, helps the user to identify important information more efficiently. \textcolor{black}{Prior work has shown that crowd-sourced content sorting and recommendation mechanisms have better performance than automated algorithms, and they are more effective in reducing misinformation \cite{crowdfake}.} 

\textit{\textbf{Displaying News by Location.}} \textcolor{black}{We found that during the crisis, people were more likely to focus on local news than they had been previously. Having the option to filter local news, or display the news on a geographic map would support this \textit{informational need}. This design would have the added benefit of helping people efficiently check for news of relevance to their family and friends, which was of great importance to our study participants.} Five participants mentioned checking the news in the locations of their family and friends. P11 wished there was a way to ``\textit{specify regions of the countries that I would care about most}.''  Moreover, when combined with sorting functions, this design enables them to find important local news more easily. 

\textit{\textbf{Distinguishing Opinion Pieces from Factual Articles.}} \textcolor{black}{Of the \textit{informational needs} raised by participants, \textit{verifying} was the most prevalent (mentioned by 19 people). Providing credibility clues could help the reader determine whether to read an article, and how to digest the information. Furthermore, the \textit{coping} need suggests an opinion tag could also help readers avoid potentially negative sentiments.} Eight participants mentioned that they would like an indicator to tell them whether a news piece is opinion or fact, such as a tag after the title. They found most opinions and quotes unhelpful for making decisions. On the other hand, P6 enjoyed reading opinion pieces to understand ``\textit{how other people might have viewed an incident},'' which helped him to form his own opinion. 

 
\subsection{Organizing News in One Place}
Participants said that news aggregation and categorization not only improves efficiency, it also helped them avoid news they did not want to see. Many participants suggested designs for improving categorization, as illustrated below. These supported users' needs to \textit{find} and \textit{verify} information, and their emotional need to \textit{contain} it.

\textit{\textbf{Creating a Hub for Crisis-related News.}} \textcolor{black}{During a crisis, the deluge of crisis-related news can overwhelm users. To help with the \textit{containing} and \textit{coping} needs expressed by participants, a news hub unifying all related resources could organize the consumption experience help users regain a sense of control over what they choose to engage with.} Three participants mentioned that they wanted to maintain a personal space when using social media, because they all experienced being exposed to an enormous number of COVID-related posts unexpectedly. Creating a hub for crisis-related news offers the user a sense of control over what they consume instead of feeling overwhelmed. P2 said, ``\textit{ I think a lot of people are just trying to focus on other things right now and actually don't want more news directly put in their face},'' and she explained that she appreciated the daily briefings from the New York Times, because they offered a dedicated section for COVID news, as well as news on other topics. \textcolor{black}{The design idea has already been adopted by many social media and news websites, where COVID-19 hubs were created in their pages. } 

\textit{\textbf{Aggregating News on the Same Topic.}} \textcolor{black}{When users described their struggles \textit{finding} information, they often mentioned encountering repetitive content across different news sources. Aggregating news on a common topic has the potential to both improve organization and also increase diversity of coverage, and aggregation has been shown to lead to more balanced coverage in other contexts \cite{Design2021trust} .} Several participants used Google News which aggregated information from different sources together (\textcolor{black}{where similar coverage is aggregated under one headline, with an expander giving the user the option to click for ``full coverage'' to reveal more reports}), which they found to be extremely helpful. 

\subsection{Providing Context and Resources}
Participants said that providing related context and resources improves users' understanding of an event, and limited contents can be misleading or not useful at all. Below are examples of participant designs that expand on the context of a news article and tie in related resources. Collectively, these design address users' needs to \textit{understand} and \textit{verify} content, and to help them \textit{cope} with it.

\textit{\textbf{Providing Related Context.}} \textcolor{black}{News information on social media is usually short and concise due to character limitation and the short attention of users.  Adding contextual information could support the \textit{understanding} need:} all participants found a news piece useful and understandable when it attached related factual reports, links or official statistics. \textcolor{black}{Consuming a single piece without knowing its background can lead to bias, as participant mentioned when they raised their need to \textit{verify} information.} For example, P10 mentioned reading an article claiming that a Belgian study had found walking could spread the virus, but the study itself did not draw the conclusion. Adding the original research article could prevent the results being misrepresented or misinterpreted. 

\textit{\textbf{Providing Actionable Resources.}} For the \textit{coping} need, Many participants described feeling burned out by consuming so much negative news. Constructive journalism \cite{Gyldensted2011InnovatingNJ} offers a pathway to address this problem: instead of only focusing on exposing the issues, the news could also offer resources and actions a reader can take. For example, P18 used an app showing case statistics in her local area that also allowed her to submit information by filling out a survey. Making this contribution was rewarding and left her feeling good about contributing to the community.


\textit{\textbf{Adding Summaries for Long Articles.}} As content has grown increasingly abundant during a crisis, attention becomes the limiting factor in the consumption of information. A long article (or video) might contain valuable resources, but it might take too much time to fully digest the piece (it might also require certain expertise to digest). Providing a gist to long articles or videos addresses the \textit{understanding} need: it helps the readers extract important takeaways, without hurting their ability to dive into the details. Ten participants mentioned that most of the time they just scan the headlines of the news without clicking to read articles, and would skip paragraphs that were too long. Three of them suggested that an article should provide a summary or overview section at the beginning to help the reader grasp the core content easily.  

\subsection{Creating a Positive News Experience.}
Participants also surfaced design ideas that they thought would help them keep a positive outlook in the face of so much negative news. These interventions supported people in meeting their needs to \textit{contain} crisis news and to \textit{cope} with it.

\textit{\textbf{Consuming Non-Crisis News.}} \textcolor{black}{As the \textit{containing} and \textit{coping} needs reveal, participants felt overwhelmed by the dominance of COVID-19 as a news topic,} giving them little opportunity to escape. P4 mentioned that she had to search for cute animal pictures to have a break, and that she started taking online courses just to get back to normal life. Participants suggested intentionally providing diverse content in the news feed, to give people the opportunity to periodically direct their attention to other topics.  

\textit{\textbf{Showing Sentiment Clues.}} The negativity of COVID-related news caused many participants to feel helpless and anxious, \textcolor{black}{as the \textit{coping} need shows.} Participants mentioned searching for positive information ``\textit{on purpose}'': P6 found himself watching more ``\textit{heartwarming}'' stories, which he had not previously done. Participants came up with several designs ideas to help them find more positive news: P10 proposed a ``\textit{good news section}'' when she found it was harder to find ``\textit{light-hearted bits}''; P18 suggested adding emojis as sentiment tags for each article to help people decide whether to engage with it or not. \textcolor{black}{For example, foxnews.com has a ``Good News'' section, and the Citizen\footnote{https://citizen.com/} app, which sends location-based alerts and updates, will regularly display positive news as a counterweight to the overall negative sentiment. Many platforms enable users to react with emojis.}

\textit{\textbf{Tracking Consumption Time.}} \textcolor{black}{Self-monitoring tools for tracking device use generally have shown to be helpful for promoting digital wellbeing \cite{Hiniker2016my, cho2021ref}. Similar tools, such as time tracking, reading status (for example, how many articles the user has read) can be embedded in interactive news platforms to raise readers' awareness and support their need to  \textit{contain} the experience and reduce endless scrolling).} P5 proposed time tracking on media to raise her awareness of her news consumption and help her spend more time in ``\textit{the real world}.'' P16 also mentioned setting a timer to remind himself to stop reading tweets. In diary responses, we also found that participants generally felt more positive when they decreased their news consumption time.  


\subsection{Making Statistics Interactive and Self-explanatory}
All participants preferred tracking the progress of the pandemic via numbers and visualizations, such as using websites \cite{noauthor_global_nodate, noauthor_covid-19_nodate} to track case numbers. Interactive maps were the most preferred visualization, because participants could check on the reported statistics using different scales, for example, switching from information about a city to information about a nation. Participants also wanted designs that provided explanations alongside statistics. For example, when P3 was viewing a prediction graph from the New York Times, she did not understand the prediction model. She felt that adding an explanation about how the data was collected and how to interpret it would improve her understanding. 

\subsection{Bringing Social Factors into Digital News}
Other designs brought social interactions into the digital news consumption experience to promote the feelings connectedness. In addition to meeting users' needs for \textit{connection}, these interventions also supported users in \textit{finding} and \textit{validating} information.

\textit{\textbf{Viewing Friends' Content.}} Many participants said they get information from their families and friends, and they trusted a news article more if it was recommended by someone close to them. \textcolor{black}{Showing friends' news-consumption activity could help address the \textit{finding} and \textit{connecting} needs. This kind of ``friendsourcing'' \cite{friendsourcing} activity personalizes the user's newsfeed, by eliciting pieces that one's friends have interacted with.} P18 brought up a feature that displayed how many friends had read an article on the app \textit{WeChat}, and commented that seeing the number was really useful because it helped her to decide whether article might be important. P6 also suggested showing the articles read by her friends, as it could help identify common topics with her friends, making her feel connected. \textcolor{black}{One example of this idea is the ``Top Stories'' section of Wechat, where the user can find articles that their friends have read and recommended.}

\textit{\textbf{Recommend Function.}} \textcolor{black}{Another strategy to interact with friends through reading the news is to ``passively'' recommend, rather than ``actively'' share a news piece. As mentioned in the \textit{connecting} need, sharing news to other people might result in a negative response,} either because they have different points of view, or they already consumed the same information. P12 suggested having a ``recommend function'' to lower the bar of sharing in a less proactive way:

\begin{displayquote}
\textit{It might be nice to include a recommended by friends article section in the Facebook COVID19 Information Center. Rather than a friend putting this on their wall (where everyone sees it in their feed), this would allow one to recommend articles for those who are interested in the topic. --- P12}
\end{displayquote}

\textit{\textbf{Voting Mechanism on Comments.}} \textcolor{black}{
Many information systems use user comments to provide complementary information to the main content \cite{revamp}, and some provide visualization tools to support digesting comments \cite{rischandrepke2021comex, Wikum}.} While most participants did not comment on news, they did view the comments as an information source. Compared to the official contents, comments sometimes ``\textit{provide other people's points of view with their personal experiences}'' (P2). To better utilize the comment section, many participants proposed a Reddit-like voting mechanism: insightful comments would be up-voted and stay on top; rude ones would be down-voted and hidden or collapsed.


\subsection{An Imaginary News App}
\label{sec:imaginary}
In order to gather participants' feedback on the design ideas, we sketched a low-fidelity news platform prototype, combining the interface related designs that could be reflected in the sketch (such as \textit{tracking consumption time}). The interface and explanation of the prototype is presented in Appendix \ref{sec:sketch}. The prototype was presented to the participants in the exit interview: 17 of the 22 participants took part in the exit interview (12 women, 5 men)\footnote{P4, P6, P14, P15, P22 did not participate in exit interviews}.

Most designs received positive feedback from all participants, including all features from \textit{Customization and Filtering} and \textit{Organizing News in One Place}. For the filter function, many participants mentioned the helpfulness of the location option: P12 commented that she would use the feature ``\textit{because I can find if something is directly impacting me locally in my town}.'' P19 would like to see an option to filter by political leaning, such as showing news from left-leaning versus right-leaning sources.

However, attitudes toward the social feature \textit{Showing Sentiment Clues} were split. In the interface, the sentiment clues of a news piece was displayed via emoji reactions. Although some participants found the sentiment cue to be an interesting concept and reported they might use it, several people raised concerns about the objectiveness and the effect of emotion manipulation. P10 mentioned that different user groups could have different reactions to an article, and for different users, the sentiment of the news could vary. Hence a single emoji might not be a meaningful way to reflect an article's sentiment. P19 brought up an issue of emotion manipulation, saying that viewing the reaction before reading the news could introduce bias.

Overall, participants offered positive feedback on all the design ideas and expressed that they would like to use those features to help improve their news consumption experience.

\section{Discussion}
We found that the COVID-19 crisis changed what users need from digital news platforms. Their information-seeking became more action-oriented and personal, as they sought out news with an end-goal of informing their own decisions. Participants encountered a new emotional cost to reading the news, as it described a high-stakes, personally relevant crisis they were actively living through. Although they found themselves struggling to cope with the burden of facing the news, they also felt an obligation to stay informed, leading to a tension wherein users were both hungry for and burdened by new information. 

Thus, participants' \textit{informational needs} and \textit{emotional needs} both became highly relevant to their use of digital news platforms. Under ordinary circumstances, the purpose of a digital news platform might be primarily to inform the public~\cite{wiki_news_2021}, but 
during the COVID-19 crisis, participants wanted digital news platforms that not only informed them but also, 1) empowered them to take action, and 2) helped them cultivate resilience to the difficult emotional work of staying informed. Achieving these aims entails prioritizing both informational needs and emotional needs as first-class design considerations. 

Participants' feedback further illustrates a concrete set of ways in which designers of digital news platforms can create experiences that address both types of needs. Here, we outline this design agenda, as surfaced by users' reactions to both existing systems and novel design ideas. Our results suggest that these design approaches can support users' interest in: finding news that informs their decision-making, avoiding unnecessary emotional costs, and bounding engagement with the news.

\subsection{Designing for Action-Oriented Information Seeking}
Participants' perspectives suggest several ways in which designers can support users' informational needs during times of crisis. First, we saw that participants' goal in reading the news shifted from an abstract interest in ``being informed'' to a concrete need to take action. As the crisis introduced great uncertainty and new risk, participants became dependent on the news for information to make vital decisions. They began to seek out more content specific to their local area, and they sought news that coupled descriptive information with actionable resources. They were frustrated when they spent time on articles that turned out to be opinion pieces, and they expressed broad exasperation with the vast sea of articles about COVID-19 they had to sift through to find the small subset that could inform their decision-making.

Collectively, these themes point to a need for crisis-sensitive news platforms to support an action-oriented approach to news consumption, that assumes users want to act on what they learn. Participants suggest designers move toward this goal by adding \textbf{customization and filters} to segment the massive landscape of news and \textbf{aggregation} to pre-organize information into subtopics. 
Although we see potential benefits to these proposed designs, complex questions remain about how to implement them and how to mitigate problematic side effects, should they arise.
For example, deciding how to support filtering, aggregation, and segmentation is no easy task, and algorithmic decisions organizing content have the potential to come with negative consequences~\cite{bias2018baeza}. An enormous amount of literature continues to examine the echo chambers and filter bubbles that arise from personalization (e.g.,~\cite{garimella2018political}), complicating participants' recommendation to aggregate similar content and filter for what is personally relevant. 
And although participants expressed interest in designs that allow users to vote on others' comments, prior research reports that human-interest and opinion content were more likely to be upvoted than news reports \cite{upvoting2014leavitt}. These examples all illustrate the need for nuance in implementing designs for this space. For example, one alternative could be dividing the general voting into several metrics, such as informativeness, usefulness, interestingness, to reflect the value of a comment in different perspectives.

Finally, participants' need to take action led them to a three-stage process of $finding \rightarrow understanding \rightarrow verifying$ information. Although each of these stages will be of relevance to news consumption in ordinary times, in moments of crisis people will need to be successful in each of these stages to make decisions with confidence. Designers of digital news platforms can use this model to interrogate their interface, examining how to design for each of these distinct phases. For example, designers might organize content by topic to improve users' ability to \textit{find} information, provide summary bubbles to help users \textit{understand} important take-aways, and add automated confidence scores to show agreement across diverse sources to help users \textit{verify} whether these take-aways are reliable. A large body of prior work offers designs that support each of these goals. For example, showing multiple sources on a single topic can help users find the diverse set of content they are looking for \cite{Perez2020counterweight}, providing summaries and statistical visualizations can reduce readers' cognitive load and improve understanding \cite{hohman2020communicating}, and priming users with specific strategies can improve their ability to spot misinformation and verify the credibility of an article~\cite{Guess15536}.
Designers can provide comprehensive support across this pipeline by drawing on these and other evidence-based techniques for each of the stages we identify. 


\subsection{Designing for Emotionally Resilient News Consumption}
When participants' lives were turned upside down by the COVID-19 crisis, they suddenly found themselves in a double bind: checking the news to stay informed suddenly became both far more \textit{necessary} but also far more \textit{emotionally taxing}. \textcolor{black}{This challenge is reflected in their shift from \textit{seeking} news to \textit{sustatining} their ability to endure the process of seeking news.} Many participants felt that they could not afford to shut out unpleasant news, because of its potential to be of direct relevance to their personal welfare or that of their loved ones. And yet, they often felt that engaging with digital news platforms took an unnecessary toll at a time when they were already under acute strain. A minority of participants found that they were unable to endure this emotional burden and \textit{did} choose to shut out the news, but this came with the cost of being under-informed at a time when being informed mattered a great deal. Many participants said they felt trapped in their own endless doomscrolling (i.e., continuing to browse negative news and feeling unable to stop), unable to look away even as digital news environments left them feeling helpless, confused, and overwhelmed. Prior work has shown that information overload can compromise mental health \cite{Josephine2018Toomuch, PENTINA2014211}, even in ordinary circumstances. In times of crisis, the potential for information overload is even greater, and our participants described feeling that COVID news is ``\textit{everywhere}.'' Simultaneously, collective mental health is compromised by crisis itself~\cite{Suresh2015disaster, green1996traumatic}, creating challenging conditions for maintaining mental well-being and strong justification for designers to attend to these needs. 

Reading about a personally relevant, actively unfolding crisis is certain to take an emotional toll that is beyond the control the designer. But participants explained that design decisions could either help users manage this burden or needlessly exacerbate it. First, designers can work to ensure the news-consumption experience is only as emotionally taxing as is necessary to meet users' informational needs. This might mean, for example, refraining from manufacturing anxiety with inflammatory headlines or viral content. Second, designers can optimize for the amount of control the user has over engaging with emotional news. \textcolor{black}{As users' design ideas suggest,} this might mean including upfront summary and sentiment details or allowing users to organize and filter content by its emotional gravity. \textcolor{black}{The needs that users expressed arose from elements of the COVID-19 pandemic that are common across many crises. A deluge of information~\cite{Hiltz2013DealingWI}, content of questionable authenticity~\cite{Sutton10twitteringtennessee}, feelings of isolation~\cite{family2013cao}, and overwhelmingly negative sentiment~\cite{Beigi2016} have all marred people's experiences living through past disasters. This suggests that platform features that participants say would sustain them as they seek to understand the COVID-19 pandemic would sustain them in other crisis contexts as well.}


Finally, \textcolor{black}{both the two-stage model of users' needs} and participants' distinct emotional needs to \textit{cope} with crisis content, \textit{contain} it, and \textit{connect} with others around it may offer useful structure to the design process. Platforms can be both created and evaluated with consideration for interface elements that support each of these distinct needs. For example, resilience theory outlines many concrete ways in which people can develop more sophisticated and successful meta-cognitive skills for coping with adversity, uncertainty, and trauma~\cite{Cheong2020what}. Similarly, mindfulness practices teach evidence-based techniques for cultivating focused attention that can enable an individual to set aside a specific concern~\cite{Brown03thebenefits, kabat2009full}. And post-crisis coping theory \cite{Hobfoll2007five} documents the importance of self- and community-connectedness in times of crisis, suggesting specific social mechanisms designers might incorporate into digital news platforms. 


\subsection{Designing for Bounded Engagement}
One of the most commonly suggested and well-received design concepts was a time-tracker surfacing to the user how much time they had spent checking the news. This was consistent with the fact that many participants said they struggled to strike a balance between ``\textit{getting enough information}'' and ``\textit{consuming too much news.}'' For many people, news-checking is a compulsive habit and source of self-frustration, even under ordinary circumstances~\cite{Ko2015NUGU, american_psychological_association_stress_2020, Lee2014hooked, Irene2015checking}. Participants told us this struggle was compounded by the crisis, as they had good reason to spend time with the news, engaging in repetitive checking habits that are known to be habit-forming~\cite{Oulasvirta2012habits}.



The monetization of users' attention~\cite{mintzer_paying_2021}, poses challenges to designing to support bounded engagement. News media platforms profit from attention and thus also use it to define their success metrics~\cite{merja2020paying}. If endless scrolling--driven by feelings of dread--increases advertising impressions and time-on-task, products are likely to optimize for it. This puts emotionally sensitive designs that encourage bounded attention at odds with many companies' profit models. However, some prior work has found that engaging with the attention economy increases traffic but not revenue to news media companies~\cite{an2018merja}. And other work has shown that people dislike and often abandon experiences that do not respect their attention~\cite{Tran2019modeling}. These and other studies (e.g.,~\cite{luo2020emotional}) suggest there is potential for news platforms that help users manage the emotional toll of the news to disrupt the status quo. 

However, there is also good reason to question whether companies creating news media platforms are able or willing to self-regulate their monetization of user attention. Public opinion increasingly favors  greater regulation and oversight from government~\cite{breman_views_2021}, which may be necessary to achieving broad change.






\subsection{Limitations}
While the research contains three stages which spanned over two weeks, the timeline spanned the first peak of the pandemic in the United States and did not capture the long-term news consumption habits. We also focused mostly on participants' digital news behaviors, thus did not ask about their engagement with other forms of news such as newspapers and radios. \textcolor{black}{All participants were in the U.S., and design preferences vary across cultures and regions~\cite{ddesigncultural}, as does media infrastructure, government oversight, and disaster response. However, users struggle with challenges like information overload, a sense of technology dependence, and overwhelming negativity in news articles in many countries (e.g.~\cite{ISLAM2020100311, fromdoomscrolling, filtering2007}), suggesting the design insights we share here are likely to have some degree of cross-cultural transferability.} In this study, we refer to the term ``crisis'' as the COVID-19 pandemic event, where all participants were located in the United States and in cities with ``stay-at-home'' orders. Prior work shows that the parameters of crisis events can vary dramatically, and news consumption habits may differ accordingly. The diary study prompting the participants to reflect may have affected their daily consumption behaviors. Finally, participants' evaluation of design sketches were speculative, and participants had no opportunity to experience them directly.

\section{Conclusion}
In this paper, we investigated how \textcolor{black}{Americans} consumed digital news during the COVID-19 pandemic and how digital news platforms can support users during times of crisis. Through the interviews, a two-week diary study, and an analysis of participant design ideas, we found \textcolor{black}{that participants moved through two different stages: an initial \textit{seeking} stage followed by a \textit{sustaining} stage. The \textit{seeking} stage was characterized by: 1) an initial spike in news consumption, 2) increases in the variety of news sources people sought out, 3) increases in consumption of local news, and 4) a shift away from negative patterns of engagement to more altruistic ones. During this stage, users' primary needs were to \textit{find}, \textit{understand}, and \textit{verify} information, all of which were systematically supported by some designs (like filters, map-based aggregators, and interactive visualizations) but not others.}

\textcolor{black}{The intensity of the \textit{seeking} stage left participants feeling overwhelmed and soon gave way to a second \textit{sustaining} stage, characterized by deliberate boundary-setting, and an active pursuit of positive and actionable news items. This stage was driven by a common set of emotional needs (specifically, to \textit{contain} crisis news, \textit{cope} with crisis news, and \textit{connect} with others around crisis news). This model provides structure for creating platforms that present news and information in times of crisis. We contribute this organizing scheme along with user-generated design examples aligned with each of its categories.}

\begin{acks}
\end{acks}

\bibliographystyle{ACM-Reference-Format}
\bibliography{refs}
\newpage
\appendix

\section{Demographics of Participants}
\label{sec:demo}
\begin{table*}[h]
\centering
\caption{Demographics of the participants in the study. Gender and ethnicity were self-reported and participants had the option to not disclose the information.}
\small
\begin{tabular}{lllllll}
\toprule
\textbf{Id}  & \textbf{Age} & \textbf{Gender} & \textbf{Ethnicity}                                                      & Occupation                                                                & \textbf{City}                                                      & \textbf{\begin{tabular}[c]{@{}l@{}}Devices use on \\ a regular basis\end{tabular}} \\ \hline
p1  & 27  & Woman & \begin{tabular}[c]{@{}l@{}}Black or \\ African \\ American\end{tabular} & Tech Supporter                                                              & New York, NY                                                       & Smartphone, Computer                                                               \\
p2  & 39  & Woman & White                                                                   & Server                                                                    & Astoria, NY                                                        & Smartphone                                                                         \\
p3  & 22  & Woman & White                                                                   & Student                                                                   & New York, NY                                                       & \begin{tabular}[c]{@{}l@{}}Smartphone, Computer, \\ TV/Radio\end{tabular}          \\
p4  & 36  & Woman & Asian                                                                   & Property Manager                                                          & Brooklyn, NY                                                       & \begin{tabular}[c]{@{}l@{}}Smartphone, Tablet, \\ Computer, TV/Radio\end{tabular}  \\
p5  & 27  & Woman & White                                                                   & \begin{tabular}[c]{@{}l@{}}Product Information \\ Specialist\end{tabular} & Detroit, MI                                                        & \begin{tabular}[c]{@{}l@{}}Smartphone, Computer, \\ TV/Radio\end{tabular}          \\
p6  & 38  & Woman & \begin{tabular}[c]{@{}l@{}}Black or \\ African \\ American\end{tabular} & Job Seeker                                                                & Seattle, WA                                                        & \begin{tabular}[c]{@{}l@{}}Smartphone, Tablet, \\ Computer\end{tabular}            \\
p7  & 30  & Man   & White                                                                   & Consultant                                                                & New York, NY                                                       & \begin{tabular}[c]{@{}l@{}}Smartphone, Tablet, \\ Computer, TV/Radio\end{tabular}  \\
p8  & 34  & Man   & White                                                                   & Stay-at-home Dad                                                          & \begin{tabular}[c]{@{}l@{}}Grosse Pointe \\ Woods, MI\end{tabular} & \begin{tabular}[c]{@{}l@{}}Smartphone, Tablet, \\ Computer\end{tabular}            \\
p9  & 28  & Woman & White                                                                   & Student                                                                   & Beachwood, NJ                                                      & \begin{tabular}[c]{@{}l@{}}Smartphone, Tablet, \\ Computer\end{tabular}            \\
p10 & 36  & Woman & White                                                                   & Attorney                                                                  & Bound Brook, NJ                                                    & Smartphone, Computer                                                               \\
p11 & 40  & Woman & White                                                                   & Computer Scientist                                                        & Bellevue, WA                                                       & \begin{tabular}[c]{@{}l@{}}Smartphone, Tablet, \\ Computer\end{tabular}            \\
p12 & 33  & Woman & White                                                                   & Teacher                                                                   & West Orange, NJ                                                    & \begin{tabular}[c]{@{}l@{}}Smartphone, Computer, \\ TV/Radio\end{tabular}          \\
p13 & 67  & Man   & White                                                                   & Educator                                                                  & Seattle, WA                                                        & \begin{tabular}[c]{@{}l@{}}Smartphone, Tablet, \\ Computer, TV/Radio\end{tabular}  \\
p14 & 20  & Man   & Asian                                                                   & Student                                                                   & Cupertino, CA                                                      & Smartphone, Computer                                                               \\
p15 & 22  & Man   & White                                                                   & Sales  Person                                                                    & New York, NY                                                       & \begin{tabular}[c]{@{}l@{}}Smartphone, Computer, \\ TV/Radio\end{tabular}          \\
p16 & 42  & Man   & White                                                                   & Web Developer                                                             & Garfield, NJ                                                       & \begin{tabular}[c]{@{}l@{}}Smartphone, Tablet, \\ Computer, TV/Radio\end{tabular}  \\
p17 & 65  & Woman & White                                                                   & Retired Editor                                                            & \begin{tabular}[c]{@{}l@{}}Fort Washington, \\ PA\end{tabular}     & \begin{tabular}[c]{@{}l@{}}Smartphone, Computer, \\ TV/Radio\end{tabular}          \\
p18 & 26  & Woman & Asian                                                                   & UX Designer                                                               & Philadelphia, PA                                                   & \begin{tabular}[c]{@{}l@{}}Smartphone, Tablet, \\ Computer\end{tabular}            \\
p19 & 20  & Man   & White                                                                   & Student                                                                   & Seattle, WA                                                        & Smartphone, Computer                                                               \\
p20 & 21  & Woman & \begin{tabular}[c]{@{}l@{}}Black or \\ African \\ American\end{tabular} & Student                                                        & Seattle, WA                                                        & Smartphone, Computer                                                               \\
p21 & 24  & Man   & Asian                                                                   & Student                                                                   & Seattle, WA                                                        & Smartphone, Computer                                                               \\
p22 & 20  & Woman & Asian                                                                   & Student                                                                   & Cupertino, CA                                                      & Smartphone, Computer                                                               \\ \bottomrule
\end{tabular}
\end{table*}

\section{Entering Interview Script}
\label{lb:interviewscript}
\subsection{News and Information Before COVID}

\begin{itemize}
    \item What smartphone system do you use? Besides your phone, what other forms of technology do you use in your daily life?
    \item Generally speaking, how much time do you spend on technology products?
    \item And what tasks/applications do you spend time on? 
    \item Before the COVID-19 pandemic, how frequently did you read news? 
    \item Did you have a regular habit of checking the news?
    \begin{itemize}
        \item \textit{Can you describe what that was like?}
    \end{itemize}
    \item Where did you get news from? 
    \item What kind of news are you interested?
    \item When reading the news, did you also engage with the content actively, such as sharing it with others or commenting on it? 
    \begin{itemize}
    \item (if yes) \textit{Describe what you did or how frequently you did it?}
    \end{itemize}
    \item Thinking back before the Coronavirus outbreak, how did checking the news usually make you feel?
\end{itemize}

\subsection{News and Information During COVID}
\begin{itemize}
    \item Where have you been getting information about COVID? Is any different resources they used before COVID?
    \begin{itemize}
        \item (if yes) \textit{Why do you use those new ones?}
    \end{itemize}
    \item When do you read news or information about COVID?
    \begin{itemize}
        \item (if yes) \textit{What prompts you to check? How often do you check?}
    \end{itemize}
    \item Is there any difference between your habits engaging with the news or seeking out information right now versus before?
    \begin{itemize}
        \item (if yes) \textit{What are they? How do you feel about the change?}
        \item (if negative feeling) \textit{Is there anything that you have tried doing to help cope with (the negative feeling)?}
        \item (if yes) \textit{Why did you try that? How effective was it?}
    \end{itemize}
    \item Could you describe a typical situation when you use a tool to look a piece of information about COVID? 
    \begin{itemize}
        \item \textit{Why is it useful? }
        \item \textit{How do you feel about this tool?}
        \item \textit{What piece do you focus on more? Why? Can you describe what that was like?}
    \end{itemize}
    \item Could you provide an example of information, or a resource about the crisis that you thought was not useful?
    \begin{itemize}
        \item \textit{Why is it not useful? }
    \end{itemize}
    \item Tell me about a time when you shared a piece of information about the crisis: 
    \begin{itemize}
        \item \textit{What did you share?}  \textit{On which platform?}
        \item \textit{Why did you share that information? What motivated you to share the piece?}
    \end{itemize}
\end{itemize}

\subsection{Design Ideas}
\begin{itemize}
    \item Finally, I’d like you to demonstrate what it’s like for you when you check the news. Can you show me what you do? As you walk me through the apps or other platforms you use, I am especially interested in the features of the app or platform and how they affect your experience.
\end{itemize}

\section{Diary Survey Questions}
\label{sec:diary}
\begin{table*}[th]
\centering
\small
\caption{The diary survey questions}
\begin{tabular}{l}
\toprule
\begin{tabular}[c]{@{}l@{}}1. Approximately how much time did you spend engaging with news and information related \\     to COVID-19 today? Do you wish you had spent less, more, or the same amount of time on this?\end{tabular}                                                \\[2ex] \hline
\begin{tabular}[c]{@{}l@{}}2. Did you make any changes today to the way you engage with information related to \\     COVID-19? If so, can you describe it?\end{tabular}                                                                                                           \\[2ex] \hline
\begin{tabular}[c]{@{}l@{}}3. What apps, websites or other technologies did you use to consume or share information \\     related to COVID-19 today?\end{tabular}                                                                                                                 \\[2ex] \hline
\begin{tabular}[c]{@{}l@{}}4. Did you notice anything about the websites, apps, or other technologies you used today \\     to access information about COVID-19 that had a \textit{positive} impact on you or left you \\     with \textit{positive} feelings? Please provide details.\end{tabular} \\[3.5ex] \hline
\begin{tabular}[c]{@{}l@{}}5. Did you notice anything about the websites, apps, or other technologies you used today \\    to access information  about COVID-19 that had a \textit{negative} impact on you or left you \\    with \textit{negative} feelings? Please provide details.\end{tabular}  \\[3.5ex] \hline
6. Based on Question 4 / 5, do you have any design ideas to improve the tool you used today? \\[0.5ex] \hline
\begin{tabular}[c]{@{}l@{}}7. How do you feel about your behavior today related to checking the news \\     or engaging with other information about COVID-19?\end{tabular}                                                                                                        \\ \bottomrule
\end{tabular}
\end{table*}

\section{News Platform Sketches}
\label{sec:sketch}
\begin{figure}[h]
\centering
\includegraphics[width=\textwidth]{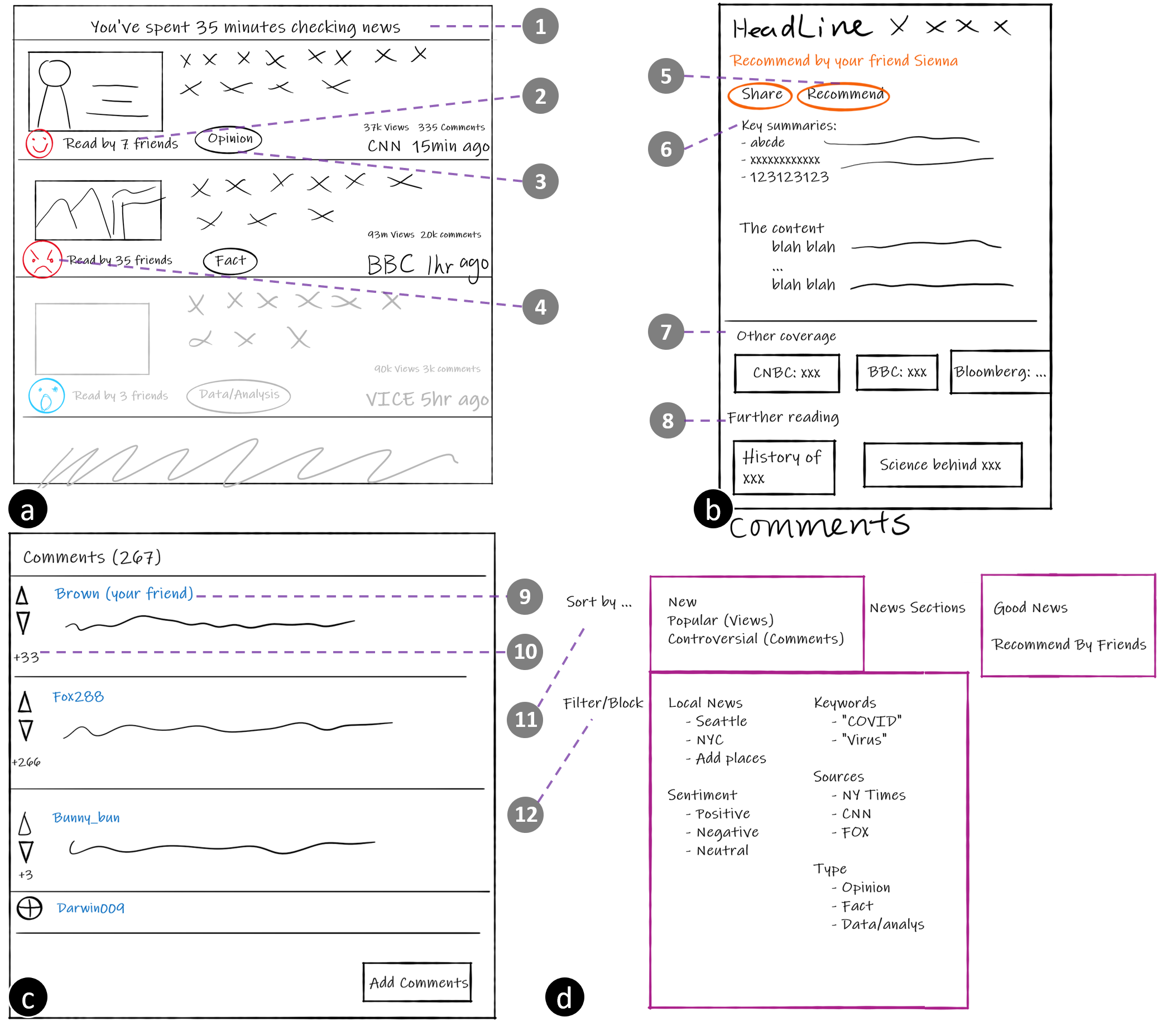}
\caption{Interface sketches of an imaginary news app: a) the feed page b) the article page c) the comment section d) the customization features. Specifically, the features are: 1. Time Tracking, 2. Read by Friends, 3. Opinion/fact/data Tag, 4. Reaction, 5. Recommendation, 6. Key Summaries, 7. Other Coverage, 8. Further Reading, 9. Friends' Comments, 10. Vote on Comments, 11. Sort Options, 12. Filter Options }
\label{fig:sketches}
\end{figure}

\textbf{\textit{Feed Page.}} The feed page is the main entrance of the news app, which includes the news feeds from different sources. On the top, there is a time tracker displaying consumption time within the app (\textit{tracking consumption time}), to raise the awareness of the user. Each piece contains a small picture and a headline. There is also an emoji indicating the sentiment of the piece (\textit{showing sentiment clues}), a label showing how many of the user's friends have read the article (\textit{display the interaction trace on a news piece}), a tag indicating which category the piece belongs to: opinion, fact, or data/analysis (\textit{providing opinion/fact clues}). Engagement statistics and basic information (such as source and time of the piece) are also included in the section. The third piece is grayed out, indicating that the user has clicked into the article before, providing information about reading progress.

\textbf{\textit{Article Page.}} After the user clicks into a news piece from the feed page, the article page is shown. There are basic elements in this page including the headline, the share button, the source, the content, and the comment section in the bottom (will be covered later). Beyond those elements, the user can also click the recommend button to ``like'' the article in a passive way (\textit{recommend function}): people will see if their friends recommended the article when they click into the piece. At the beginning of the article, there is also a ``key summaries'' section providing the gist of the content (\textit{adding summaries for long articles}). At the end of the article, there are also two sections providing extra resources: ``other coverage'' displays the news on the same topic from other sources (\textit{aggregating news on the same topic}), and ``further reading'' includes the background stories and related resources of the article (\textit{providing related context}, \textit{providing actionable resources}). 

\textbf{\textit{Comment Section.}} The comments can be up/down voted, and are sorted by votes. Friends' comments will be shown on the top. If a comment is voted down enough times, it will be collapsed (\textit{voting mechanism on comments}).

\textbf{\textit{Customization Features.}} We also included a customization page showing the functions related to information organization such as sorting, filtering and categorizing. The feeds can be sorted by time/views/comments, and users can filter or block news based on location/sentiment (indicated by the emojis)/keyword/source/type (\textit{adding granularity to customization and filter features}). Lastly, there are two new sections in the feed that users can browse: ``good news'' and ``recommended by friends.''

\end{document}